\documentclass[12pt,cite,epsf,epsfig]{article}
\usepackage{epsfig}
\title{$B \to X_s +$ missing energy in models with large extra dimensions}
\author{Namit Mahajan\thanks{E--mail : nm@ducos.ernet.in, 
nmahajan@physics.du.ac.in}\\
	{\em Department of Physics and Astrophysics,} \\
	 {\em University of Delhi, Delhi-110 007, India.}}

\def\be{\begin{equation}}
\def\ee{\end{equation}}
\def\bea{\begin{eqnarray}}
\def\eea{\end{eqnarray}}

\begin{document}
\maketitle

\begin{abstract}
We study the neutral current flavour changing rare decay mode
$B\rightarrow X_s +$ missing energy within the framework of
theories with large extra spatial dimensions. The corresponding Standard Model
signature is $B\rightarrow X_s + \nu \bar{\nu}$. But in theories with large
extra dimensions, it is possible to have scalars and gravitons in the final
state making it quite distinct from any other scenario where there are 
no gravitons and the scalars are far too heavier than the B-meson to be
present as external particles. We give an estimate of the branching ratio
for such processes for different values of the number of extra dimensions and 
scale of the effective theory. The predicted branching ratios can be comparable with
the SM rate for a restrictive choice of the parameters.
\end{abstract} 

\begin{section}{Introduction}
The investigations of flavour changing neutral current (FCNC) transitions
of b-quark offer excellent opportunities to test the basic structure
of the underlying theory. These processes are quite sensitive to QCD
and possible long distance corrections as well as contributions from
new particles in the loops. Consequently, such processes become useful tools for testing
new physics as well. The measurement of $B \to X_s \gamma$ (and 
subsequently the exclusive channel $B \to K^* \gamma$) by CLEO \cite{cleo}
has been used to stringently constrain the parameter space of supersymmetric
(SUSY) and various other theories (see for example \cite{const} and references
therein).
The same is true for any other FCNC process. In particular, the quark
level transition $b \to s \nu \bar{\nu}$ can turn out to be a very 
successful place for putting strong and meaningful bounds on the underlying
 theory. The Standard Model (SM) experimental signature in this case is 
the observation
of the decay process $B\to X_s +$ missing energy in the form of neutrinos.   
In any extension of the SM, decay of a b-quark to an s-quark and neutral,
 ultra light particle(s) can mimic the SM process.\\

It is well known that SM is plagued with the hierarchy problem and many
possible solutions have been proposed in the past. However, the idea that
the fundamental scale of gravitational interaction being quite distinct
from the Planck scale ($\sim 10^{19}$ GeV) and possibly
as low as ${\mathcal{O}}$(TeV) has attracted a lot of attention. In the simplest version,
as put forward by Arkani-Hamed et.al. (referred to as ADD model/scenario from
now on) \cite{nima}, the basic idea is 
the existence of n spatially compact large dimensions. The spacetime
is a direct product of the four dimensional Minkowski space and 
the compact space spanned by the n spatial extra dimensions. The SM fields
are all resticted to a 3-brane while gravity is free to propagate in the bulk.
Seen from a four dimensional point of view, it simply means that apart from
the usual massless graviton mode, there is a tower of Kaluza-Klein (KK) 
excitations from the gravity sector due to compactification. Also, if $M_{*}$
and $M_{Pl}$ denote the effective scale of gravity and the four dimensional
Planck scale respectively and if $R$ is the radius of compactification
of the extra dimensions (assuming the same radius for all the n dimensions), 
then it turns out that they are related as follows
\begin{equation}
M_{Pl}^2 \sim R^n~M_{*}^{n+2} \label{eq.1}
\end{equation} 
Therefore, for large enough values of $R$, the effective scale $M_*$ can be as
low as ${\mathcal{O}}$(TeV) and still be consistent with Eq.(\ref{eq.1}). Although 
the individual KK modes couple
to the SM fields on the 3-brane by the ordinary gravitational strength
but the presence of a large number of them makes the effective coupling
appear to be of the order of TeV$^{-1}$ rather than Planck scale inverse.
This means that gravity starts to become a strong force at TeV scales in sharp
contrast to the situation in any ordinary theory of gravity.\\

In the present study, we investigate the inclusive decay $B \to X_s +$ missing
energy in the context of the ADD scenario. In the SM or most of its extensions,
the missing energy is in the form of the neutrinos that escape without
being detected and it is only new contributions to the relevant Wilson 
coefficients that are induced by interactions beyond SM. However, there is a sharp
contrast in the case at hand. In the present scenario, there can be very light
gravitons and associated scalars (the dilatons) that can be present in the
final state. The situation is clearly distinct from SM or its usual 
extensions as in all such cases there are no spin-2 gravitons involved and
the scalars, both neutral and charged, are far too massive compared to the b-quark
to be present in the external legs. But this is possible in theories with
large compact extra dimensions. It is precisely this advantageous aspect
that we would like to exploit in the present study and see whether we
get any meaningful results.
\end{section}

\begin{section}{$b \to s + KK$ modes}
The effective Hamiltonian for $b \to s$ transition in the SM is \cite{buchalla}
\be
{\cal{H}}_{eff} = -\frac{G_F}{\sqrt{2}}V_{ts}^*V_{tb}\sum_i 
                         C_i(\mu)O_i(\mu),
\ee
with
\[
O_1 = (\bar{s}_ic_j)_{V-A}(\bar{c}_jb_i)_{V-A},
\]
\[
O_2 = (\bar{s}_ic_i)_{V-A}(\bar{c}_jb_j)_{V-A},
\]
\[
O_3 = (\bar{s}_ib_i)_{V-A}\sum_q(\bar{q}_jq_j)_{V-A},
\]
\[
O_4 = (\bar{s}_ib_j)_{V-A}\sum_q(\bar{q}_jq_i)_{V-A},
\]
\be
O_5 = (\bar{s}_ib_i)_{V-A}\sum_q(\bar{q}_jq_j)_{V+A},
\ee
\[
O_6 = (\bar{s}_ib_j)_{V-A}\sum_q(\bar{q}_jq_i)_{V+A},
\]
\[
O_7 = \frac{e}{16\pi^2}\bar{s}_i\sigma^{\mu\nu}(m_sP_L +
m_bP_R)b_iF_{\mu\nu},
\]
and
\[
O_8 = \frac{g}{16\pi^2}\bar{s}_i\sigma^{\mu\nu}(m_sP_L + m_bP_R)
           T_{ij}^ab_jG_{\mu\nu}^a.
\]
Apart from these there are semi-leptonic operators as well
\[
O(ee)_V = (\bar{s}_ib_i)_{V-A}(\bar{e}e)_V
\]
\be
O(ee)_A = (\bar{s}_ib_i)_{V-A}(\bar{e}e)_A
\ee
and finally 
\[
O(\nu\bar{\nu}) = (\bar{s}_ib_i)_{V-A}(\bar{\nu}\nu)_{V-A}
\]
It is this last operator that is responsible for the relevant decay channel in 
SM. In order to calculate any process beyond these known operators, one will 
have to either write down all possible new operators respecting the symmetries
of the low energy theory and then indirectly fix the associated coefficients
or explicitly make the calculation. We follow the latter route.\\

The coupling of the gravitons and the dilatons can be obtained from the
low energy effective action \cite{tao,giudice}. As expected, 
the gravitons couple
to the energy-momentum tensor of the SM fields and the dilaton
couples to the trace of the energy-momentum tensor. It is then
straight forward to get the Feynman rules and compute the individual
contributions. We consider the dilaton/scalar-KK emission first and comment on the
analogous graviton emission later. The diagrams contributing to the
dilaton emission process are shown in Fig.1.
It may be important to mention here that: (i) there is no momentum dependent piece
in the trace of energy-momentum tensor for the fermions and (ii) the 
dilaton-gauge boson-fermion-fermion vertex does not exist. However, both these are shown to be present
in \cite{tao}. The fact that it is indeed so is not hard to see. The trace of the energy-momentum tensor
for the fermionic plus gauge bosonic parts (including interaction term)
is nothing but 
\be
T^{\mu}_{\mu} = -m_f\bar{f}f + m_V^2A^2 \label{eq.2}
\ee
In obtaining this, one needs to consider the interacting Heisenberg field
equations rather than the free ones. The use of full interacting equations of motion
for the fermion and the gauge boson results only in the mass terms for each of them
(similar expression is obtained for the trace of the energy-momentum tensor in \cite{cheung})
and thus there is no momentum dependence for the fermions involved in the Feynman rules.
Also, the gauge boson-fermion-
fermion-dilaton vertex will never be present (although a similar vertex for
the gravitons will be there as in \cite{tao}). Thus, we differ on these points with the
rules given in \cite{tao} and it may be important to emphasize again that in obtaining
the energy-momentum tensor or its trace, nowhere has the on-shell condition been imposed. Therefore,
the result Eq.(\ref{eq.2}) is exact and true in general. Similar remarks and results were quoted \cite{nm} in the context of
a loop calculation in the Randall-Sundrum scenario, where the scalar in the theory (called the radion)
couples, at the first order, with the trace of the energy-momentum tensor of the SM fields
and is very similar
to the dilaton in the ADD scenario as far as the couplings with the SM fields are concerned.
 There are less number of diagrams contributing to the dilaton emission process as compared to the 
graviton emission process. Also, in the limit of neglecting the strange quark
mass, the diagram with dilaton being emitted from the s-quark line also
vanishes.\\

Clearly, all the diagrams are divergent and there is no underlying symmetry
like the one present in SM that ensures cancellations between the individual
diagrams to render a finite result. By naive power counting, one finds that
there are hard divergences (see Appendix) which are quadratic, quartic and possibly even 
higher in the case of graviton emission. It now brings the question 
of handling such divergences in the context of effective field
theories. It has been very strongly advocated \cite{london} that in any
sensible effective field theory all the quadratic and higher divergences
are cancelled by counter-terms arising out of the complete underlying 
high-energy theory. At the one loop level, it is thus the 
logarithmic dependence
on the cut-off only that can be extracted from the low energy effective
theory. Thus, it makes more sense to keep only the logarithmic pieces as
far as the one loop calculation within an effective theory is concerned.
We adhere to this approach in our calculations and thus use dimensional
regularization throughout and at the end replace $1 \over{\epsilon}$ by
$\ln\Bigg({\Lambda^2 \over{m_W^2}}\Bigg)$ with the identification $\Lambda = M_*$.\\

As mentioned earlier, it is expected that the presence of the scalars and
gravitons in the external legs makes the situation more interesting as new 
operators are introduced. The operators are of the form (denoting the dilaton 
and the graviton fields by $\Phi$ and $h_{\mu\nu}$ respectively)
\be 
O(dilaton) \sim (\bar{s}_i(1\pm\gamma_5)b_i) \Phi \label{eq.6}
\ee
and
\be 
O(graviton) \sim (\bar{s}_i\gamma_{\mu}{p_b}_{\nu}(1\pm\gamma_5)b_i) h^{\mu\nu} \label{eq.7} 
\ee 
(see Appendix for details)

The invariant matrix element for the quark level process $b \to s \Phi^{(n)}$
for a dilaton of mass $m_n$ is (the individual; contributions are listed in Appendix)
\bea
{\mathcal{M}}(b \to s \Phi) &=& \Bigg({\imath\over{16\pi^2}}\Bigg)
{G_F\omega\kappa\over{\sqrt{2}}}\Bigg({1\over{\epsilon}}\Bigg)
\Bigg(\sum_i \xi_i m_i^2\Bigg)\Bigg({m_b\over{3}}\Bigg) \\ \nonumber
&&\Bigg[18\Bigg({m_s^2 \over{m_b^2-m_s^2}}\Bigg) - 1\Bigg]\bar{s}(1+\gamma_5)b \label{eq.3}
\eea
where $\xi_i$ is the CKM factor, $m_i$ is the mass of the up-type quark in the 
loop. We have neglected ${\mathcal{O}}({m_s^2\over{m_W^2}})$ and higher terms
 in obtaining this expression. Using this matrix amplitude it is now trivial 
to compute the decay rate
for the emission of a single dilaton of mass $m_n$. For the inclusive process,
it suffices to use the quark level amplitude to get the decay rate.
 One should sum over the
tower of these dilatons till the b-quark mass scale. Following the summing 
techniques illustrated in \cite{tao}, we get the following expression for the
decay rate 
\bea
\Gamma(B\to X_s \Phi) &=& \Bigg({1\over{16\pi^2}}\Bigg)^2
{G_F^2\over{2}}\Bigg[\ln\Bigg({M_*^2\over{m_W^2}}\Bigg)\Bigg]^2
\Bigg(\sum_i \xi_i m_i^2\Bigg)^2\Bigg({4m_b\over{27}}\Bigg) \\ \nonumber
&\times&\Bigg({1\over{n+2}}\Bigg)
\Bigg[{1\over{n}} + {1\over{n+4}} - {2\over{n+2}}\Bigg]
\Bigg({m_b\over{M_*}}\Bigg)^{n+2}
\eea 
where we have made the replacement 
\[
{1\over{\epsilon}} \longrightarrow \ln\Bigg({M_*^2\over{m_W^2}}\Bigg)
\]
The expression for the graviton emission rate will look similar in its final
form (see Appendix) but the rate itself is expected to be higher because of two
 reasons. Firstly, there are two extra diagrams (with the graviton being
attached to either of the ferrmion-fermion-gauge boson vertex) and secondly,
the graviton coupling does not have the $\omega$ factor suppression present
in the dilaton coupling. Thus, the rate is expected to be
\be
\Gamma(B\to X_s G) \sim {\mathcal{F}}~\Gamma(B\to X_s \Phi) \label{eq.4}
\ee
where ${\mathcal{F}}$ is a numerical factor expected to be of the order $10$ or higher.
However, it is instructive to go through the complete steps for the graviton emission
process as well. The procedure is completely in analogy with the dilaton emission. The total
rate is obtained by summing over various massive modes in the final state. Using the expression
for the rate corresponding to a single graviton of mass $m_n$, the summation essentially involves integrating
over the tower of the massive modes till the b-mass scale. When this is done, it is found that there is a divergent
behaviour for small $n$, the number of extra dimensions. This is like the soft emission rates diverging for some particular 
values of $n$ and arises due to the structure of the integrand.
We thus quote the values only for $n ~\ge~ 4$ (in particular $n~=~4,~6$). In fact, the limit $n\to 4$
has also to be taken with care and in some sense in the spirit of analytic continuation or something similar.     
\end{section}

\begin{section}{Results}
The total contribution to the process $B \rightarrow X_s + KK$ modes is
\[
\Gamma(B\to X_s \Phi) + \Gamma(B\to X_s G)
\]
Below we quote the branching fraction for the dilaton mode for different
values of $n$ and $M_*$.
\begin{table}[ht]
    \caption{${\it{Br}}(B\to X_s + \Phi)$}
    \begin{center}

    \begin{tabular}{|c|c|c|}
	\hline
	$n$ & $M_*$ (TeV) & Branching ratio \\
		\hline
	2	&1 	& $1.28 \times 10^{-3}$\\ 
       3 	& 1     & $2.25 \times 10^{-6}$ 	\\ 
       4        & 1     & $4.93 \times 10^{-9}$ \\ 
\hline
	2	&5 	& $5.50 \times 10^{-6}$\\ 
       3 	& 5     & $1.93 \times 10^{-9}$ 	\\ 
       4        & 5     & $8.45 \times 10^{-13}$ \\ 
\hline
	2	&10	& $4.69 \times 10^{-7}$\\ 
       3 	& 10    & $8.23 \times 10^{-11}$ 	\\ 
       4        & 10    & $1.80 \times 10^{-14}$ \\ 
\hline
	2	&50	& $1.33 \times 10^{-9}$\\ 
       3 	& 50    & $4.68 \times 10^{-14}$ 	\\ 
       4        & 50    & $2.05 \times 10^{-18}$ \\ 
\hline
\end{tabular}
\end{center}
\end{table}
The branching ratio for the graviton mode for $n=4$ and $n=6$ are as follows:
\begin{table}[ht]
    \caption{${\it{Br}}(B\to X_s + G)$}
    \begin{center}

    \begin{tabular}{|c|c|c|}
	\hline
	$n$ & $M_*$ (TeV) & Branching ratio \\
		\hline
	4	&1 	& $1.02 \times 10^{-4}$\\ 
         6        & 1     & $3.2025 \times 10^{-11}$ \\ 
\hline
	4	&5 	& $1.72 \times 10^{-8}$\\ 
         6        & 5    & $2.2 \times 10^{-16}$ \\ 
\hline
	4	&10 	& $3.67 \times 10^{-10}$\\ 
         6        & 10     & $1.17 \times 10^{-18}$ \\ 
\hline
	4	&50 	& $4.17 \times 10^{-14}$\\ 
         6        & 50     & $5.33 \times 10^{-24}$ \\ 
\hline
\end{tabular}
\end{center}
\end{table}

These numbers are to be compared with the SM expectation, ${\it Br}(b\to s
\nu\bar{\nu}) = 5 \times 10^{-5}$ and the experimental limits on the
same, ${\it Br}(b\to s\nu\bar{\nu})_{exp} < 6.2 \times 10^{-4}$ at $90\%$
confidence level \cite{aleph}.
From the Table 1 it is clear that for $n=2$ and $M_*=1$ TeV, the contribution
far exceeds the SM prediction and the experimental limits and should have been
observed. Moreover, one encounters some divergent results for the 
graviton emission process for smaller values of $n$ which may be looked upon as
arising due to something similar to gravi-strahlung typical to these smaller values of $n$.
We therefore cannot say anything very satisfactorily for these values unless a more careful and detailed
analysis is done. Thus we restrict ourselves to dilaton rates for these values of $n$ and don't consider graviton rates
for those values.\\

Also, evident is the fact that for higher values of $n$ and/or $M_*$, the
branching fraction is smaller. The constraints obtained from 
the collider experiments \cite{uehara} allow even smaller values of $M_*$ than quoted, 
thus allowing for the possibility of some more enhancement in the branching ratio predictions.  
 However, strong constraints from the supernova SN1987A observations \cite{cullen}
and cosmology \cite{hall} rule out such choices of $n$ and $M_*$. For $n=2$, SN1987A
constraints imply $M_* \sim50$ TeV. Even if one takes care of all the uncertainties
entering the supernova calculations, $M_*=1$ TeV seems to be far from reachable. The other
values quoted in the table are more or less allowed by these constraints but seem to be contributing
very little. However, for $n=4$ and $M_*$ greater than a few TeV, the sum of the dilaton and graviton
emission rates can be sizeable and may compete with the SM rate. The constraints coming from
other astrophysical and cosmological processes \cite{uehara, hannestad} 
like cosmic diffuse gamma ray background, early matter domination etc are even stronger than quoted above and
rule out any chances of extra dimensional contribution to the decay process being comparable to the SM rate
and thus being observable at the future B-factories. Also, given such strong constraints,
the idea of having different radii
for different compact dimensions, or atleast partially, does not seem to
be of much significance. \\

It may be useful to mention here that the mode $B\to X_s G$ can contribute to
SM process $B\to X_s \nu\bar{\nu}$ via $G\to \nu\bar{\nu}$ and the sum of
the SM and the extra contribution will form the complete matrix element.
But, in the case of the scalar mode, $B\to X_s \Phi$, there can be no cascade
decay of $\Phi$ to $\nu\bar{\nu}$. Hence, until and unless the produced
$\Phi$ decays into low mass particles and escapes as missing energy,
the energy and angular spectrum of the hadronic junk, $X_s$, will be
markedly different from the corresponding distribution observed in the case of
$B\to X_s \nu\bar{\nu}$ coming from any theory. The same is true for the 
graviton emission process provided it does not decay further into
$\nu\bar{\nu}$ or any other pair of low mass particles.
From the structure of $O(graviton)$ it is evident that the operator
corresponding to $b\to s \nu\bar{\nu}$ will have the form
\[
O(\nu\bar{\nu})_{graviton} \sim  (\bar{s}_ib_i)_{V\pm A}(\bar{\nu}\nu)_V
\]
which is pure vector current for the $\nu\bar{\nu}$ pair, a feature different
 from the SM again where the intermediate particle responsible is the 
Z-boson, making the angular distribution different.
Also observe that if instead of following the approach advocated in 
\cite{london}, we had retained quadratic and higher divergences, we
would have obtained enormously large, intolerable numbers. This again
justifies the retaining of only logarithmic terms and in turn justifies 
our use of dimensional regularization in carrying out the calculations, which 
otherwise would have to be carried out with a hard ultra-violet cut-off.\\

In conclusion, we can say that for two extra dimensions and with $M_*\ge$
 few TeV (as marginally allowed by the constraints from astrophysics and cosmology),
 the desired branching ratio can be sizeable. Other choices of n seem to be too
small to compete with the SM rate. Also, for $n>2$, the radius of compactification, $R$, becomes too
small for its effects to be probed at future mechanical experiments testing the gravitational law
at small length scales (eg. see \cite{long} for details of these experiments).
 Thus, theories with large extra dimensions
predict a competing value of the branching ratio for the process $B\to X_s + $ missing energy
(in the form of soft dilatons and gravitons) for a very restrictive choice of the parameters.
 One can thus hope that future observation of the decay mode
$B\to X_s +$ missing energy, along with the corresponding exclusive modes, will be able to
testify theories with large extra dimensions and if not completely ruled out, yield
severe constraints on the number of extra dimensions and the effective
scale of gravitational interactions in these theories.

\end{section}

\begin{section}*{Appendix}
We outline the steps involved in the evaluation of a typical loop diagram and quote the expressions
for individual matrix elements. For the sake of illustration of the calculation, we evaluate the loop diagram (b), where
the dilaton is emitted from the W-propagator. We call this matrix element ${\mathcal{M}}_{(b)}$. Label the momenta as
$p_b$, $p_s$ and $p_{\Phi}$ corresponding to b-quark, s-quark and the dilaton and denote the loop momentum variable by $k$.
The matrix element can be written as
\bea
{\mathcal{M}}_{(b)} &=& \Bigg[~~\Bigg] \int{\tilde{dk}}
\frac{[\eta^{\mu\alpha}-\frac{(k-p_{\Phi})^{\mu}(k-p_{\Phi})^{\alpha}}{m_W^2}]
[\eta^{\nu\beta}-\frac{k^{\nu}k^{\beta}}{m_W^2}]}{(k^2-m_W^2)[(k-p_{\Phi})^2-m_W^2][(p_b-k)^2-m_i^2]} \\ \nonumber
&\times& \eta_{\alpha\beta}~\bar{s}(p_s)
[\gamma_{mu}(1-\gamma_5)(\not{p_b}-\not{k}+m_i)\gamma_{\nu}(1-\gamma_5)]b(p_b)
\eea
where
\[
\Bigg[~~\Bigg] = \Bigg[-\sum_{i}\xi_i\Bigg(\frac{g}{2\sqrt{2}}\Bigg)^2 \omega\kappa m_W^2\Bigg] \,\, ;
\hskip 1.4cm {\tilde{dk}} = \frac{d^4k}{(2\pi)^4}
\]
In the above expression, $g$ is the weak coupling constant, $\xi_i$ is the relevant CKM factor,
$m_i$ is the mass of the up-type quark in the loop, $\kappa^2 = 16\pi G_N$ gives the gravitational
coupling in terms of Newton's constant $G_N$ and $\omega$ is a number that depends only on the number
of extra dimensions, n, and is given by
\[
\omega^2 = \frac{2}{3(2+n)}
\]  
Clearly, $\omega <1$ and gives a supression factor. Also, it can be seen very clearly from the above expression for
the matrix element that the $k$-integral is not logarithmically divergent by power counting but there are higher divergences
present. The reason for the presence of such hard divergences is the fact that there is no underlying symmetry,
like the one present in the SM, to ensure nice cancellations and render a finite or at the most logarithmically
divergent result. This can be attributed to the non-renormalizable nature of these interactions. The calculation
has been done in teh unitary gauge and it may seem that these power divergences are arising due to the form of the
massive gauge boson propagator. However, it should be noted that if instead one works in some other convenient gauge,
say Feynman gauge, the massive gauge boson propagators have a more well behaved high energy structure but there are
additional diagrams involving the asociated ghosts which essentially are the badly behaved longitudinal components
of these massive gauge bosons and also that their interaction vertices contain momentum dependence. Therefore, with some
effort, various terms in the calculation in some other gauge can be combined to give the same results.
This is also evident from the Feynman rules in \cite{tao} where the vertices have an additional gauge dependence. 
 But we follow the thumb rule \cite{london} that only logarithmically divergent terms be retained and 
thus use dimensional regularization (in the full spirit of an effective theory calculation)
throughout our calculation. The integral can be evaluated using standard procedure of
Feynman parameterization, shifting the variables and carrying out the integrations. We don't evaluate the finite pieces and
only retain divergent terms. After the Feynman parameterization and shifting of momentum variables, the terms that are not
multiplied by $m_i$ (or its powers) drop out due to the unitarity of the CKM matrix (the GIM mechanism).
 Retaining just the various divergent terms
in the integral and employing dimensional regularization (neglecting terms 
${\mathcal{O}}({m_s^2\over{m_W^2}})$ and higher), we arrive at the following expression for the matrix element corresponding
to the diagram (b):
\bea
{\mathcal{M}}_{(b)} &=& \Bigg(\frac{\imath}{16\pi^2}\Bigg)\Bigg[4\sum_{i}\xi_i\Bigg(\frac{g}{2\sqrt{2}}\Bigg)^2 \omega\kappa 
\frac{m_i^2}{m_W^2}\Bigg]\Bigg(\frac{1}{\epsilon}\Bigg) \\ \nonumber
&& \bar{s}(p_s)\Bigg[\frac{3}{4}m_s(1-\gamma_5) + \frac{2}{3}m_b(1+\gamma_5)\Bigg]b(p_b)
\eea 
where $m_b$ and $m_s$ are the masses of the b- and s-quark respectively. Also, using the relation 
$\frac{G_F}{\sqrt{2}}=\frac{g^2}{8m_W^2}$ we can express the result in terms of Fermi-constant $G_F$.\\

The expressions for other diagrams are as follows:
\bea
{\mathcal{M}}_{(a)} &=& \Bigg(\frac{\imath}{16\pi^2}\Bigg)\Bigg[4\sum_{i}\xi_i\Bigg(\frac{g}{2\sqrt{2}}\Bigg)^2 \omega\kappa 
\frac{m_i^2}{m_W^2}\Bigg]\Bigg(-\frac{3}{4}\Bigg)\Bigg(\frac{1}{\epsilon}\Bigg) \\ \nonumber
&& \bar{s}(p_s)\Bigg[m_s(1-\gamma_5) + m_b(1+\gamma_5)\Bigg]b(p_b)
\eea 

\bea
{\mathcal{M}}_{(c)} &=& \Bigg(\frac{\imath}{16\pi^2}\Bigg)\Bigg[4\sum_{i}\xi_i\Bigg(\frac{g}{2\sqrt{2}}\Bigg)^2 \omega\kappa 
\frac{m_i^2}{m_W^2}\Bigg]\Bigg(-\frac{3}{4}\Bigg)\Bigg(\frac{1}{\epsilon}\Bigg)\frac{m_bm_s}{(m_b^2-m_s^2)} \\ \nonumber
&& \bar{s}(p_s)\Bigg[m_b(1-\gamma_5) - m_s(1+\gamma_5)\Bigg]b(p_b)
\eea 

\bea
{\mathcal{M}}_{(d)} &=& \Bigg(\frac{\imath}{16\pi^2}\Bigg)\Bigg[4\sum_{i}\xi_i\Bigg(\frac{g}{2\sqrt{2}}\Bigg)^2 \omega\kappa 
\frac{m_i^2}{m_W^2}\Bigg]\Bigg(\frac{3}{4}\Bigg)\Bigg(\frac{1}{\epsilon}\Bigg)\frac{m_bm_s}{(m_b^2-m_s^2)} \\ \nonumber
&& \bar{s}(p_s)\Bigg[m_b(1-\gamma_5) + m_s(1+\gamma_5)\Bigg]b(p_b)
\eea 
Adding all these contributions we arrive at Eq.(\ref{eq.3}).\\

Similar to the dilaton emission process, we take a look at the graviton emission process and the structure that is obtained.
We compute the graviton emission process corresponding to diagram (a). Following the same steps as outlined
above we get for the matrix element and neglecting $m_s$
\bea
{\mathcal{M}}_{(a)}^{graviton} &=& \Bigg(\frac{\imath}{16\pi^2}\Bigg)\Bigg[-\sum_{i}\xi_i\Bigg(\frac{g}{2\sqrt{2}}\Bigg)^2 \kappa 
\frac{m_i^2}{m_W^2}\Bigg]\Bigg(\frac{1}{3}\Bigg)\Bigg(\frac{1}{\epsilon}\Bigg) \\ \nonumber
&& \bar{s}(p_s)\gamma_{\alpha}{p_b}_{\beta}(1-\gamma_5)b(p_b)~\epsilon^{\alpha\beta}
\eea 
where $\epsilon^{\alpha\beta}$ is the polarization tensor for the graviton field. Other diagrams give similar contributions.
Due to momentum dependent factors in the Feynman rules (when a graviton is involved), the divergence structure
is even worse and naive power counting shows that divergences higher than quadratic are present. We don't work in
the cut-off scheme and the use of dimensional regularization again ensures that we retain only the logarithmic divergences.
Note that there is no factor of $\omega$ appearing in the above expression 
and thus the supression due to this factor is absent in the graviton emission
process and depending on the number of extra dimensions etc., the rate for the graviton emission process can be larger than 
the dilaton emission rate by a factor of 10 or more. 
This extra multiplicative factor has been identified as ${\mathcal{F}}$ in Eq.(\ref{eq.4}).  
The graviton emission rate (for a single graviton of mass $m_n$) is 
\bea
\Gamma(B\to X_sG^{(n)}) &=& \Bigg({1\over{16\pi^2}}\Bigg)^2
{G_F^2\over{2}}\kappa^2\Bigg[\ln\Bigg({M_*^2\over{m_W^2}}\Bigg)\Bigg]^2
\Bigg(\sum_i \xi_i m_i^2\Bigg)^2\Bigg({5\over{27}}\Bigg)\Bigg({1 \over{32\pi}}\Bigg) \\ \nonumber
&\times& \Bigg(1-\frac{m_n^2}{m_b^2}\Bigg)~\Bigg[\frac{(m_b^2-m_n^2)^3(2m_b^2+3m_n^2)}{m_n^4}\Bigg]
\eea
Summing over all the states lying till the b-quark mass scale gives the final expression for graviton emission
rate. Clearly, the form for the expression is similar to that of the dilaton case and there are no factors of $\omega$ 
present. It is important to remark that from the above expressions, it is clearly evident that the oerator
structure for the dilaton or graviton emission processes is indeed the ones quoted in Eqs.(\ref{eq.6})-(\ref{eq.7}).\\
\end{section}
\vskip 1cm
{\bf{Acknowledgments}}\\ NM would like to thank the University Grants Commission,
India, for financial support.
\vskip 3cm
\begin{figure}[ht]
\vspace*{-1cm}
\centerline{
\epsfxsize=5cm\epsfysize=3.5cm
                    \epsfbox{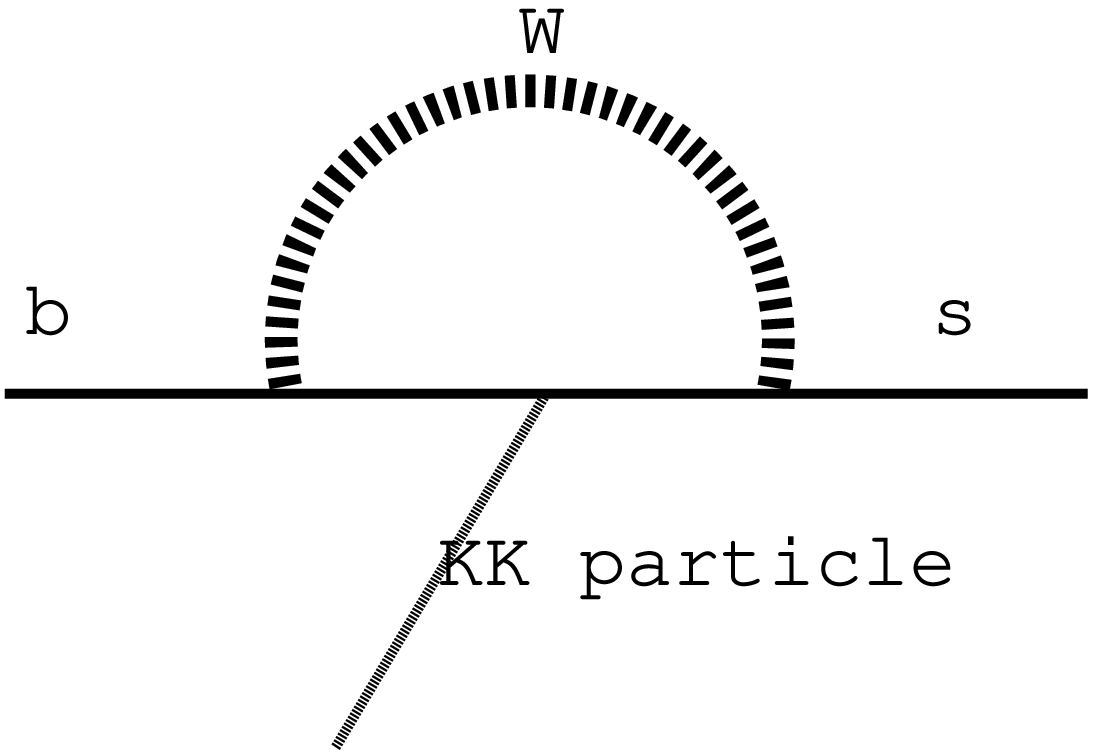}
\hskip 1.5cm
\epsfxsize=5cm\epsfysize=3.5cm
                    \epsfbox{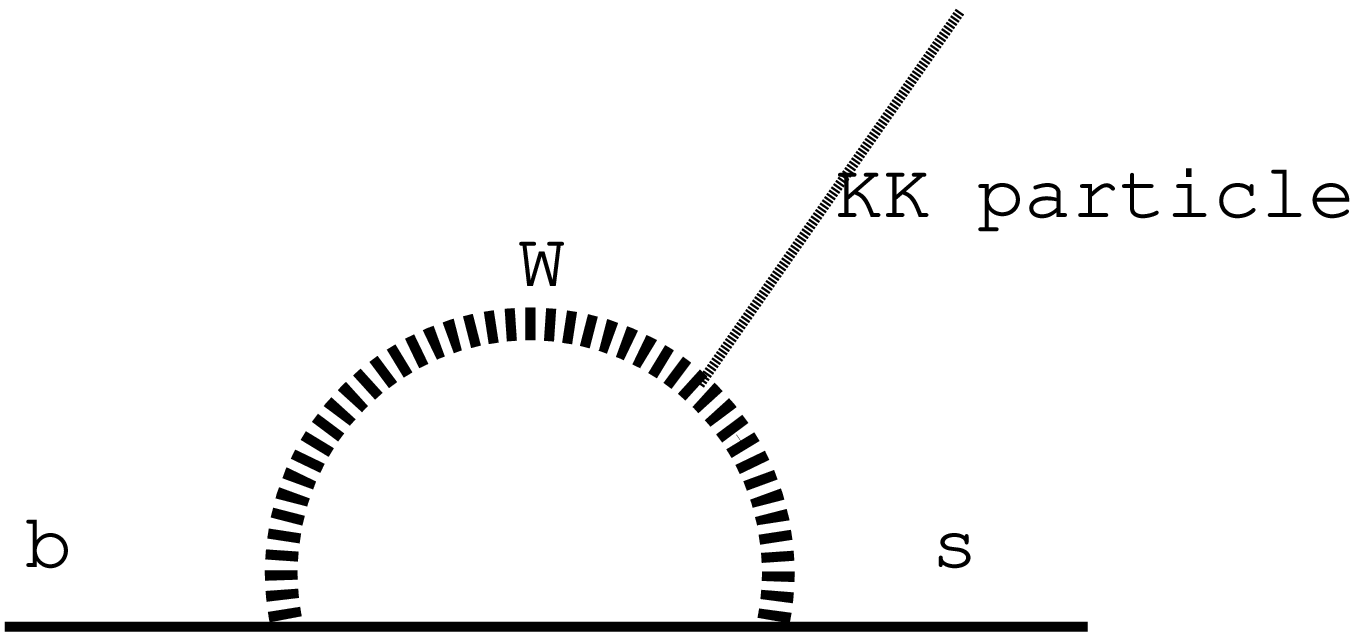}}
\vskip 1cm
\centerline{
\epsfxsize=5cm\epsfysize=3.5cm
                    \epsfbox{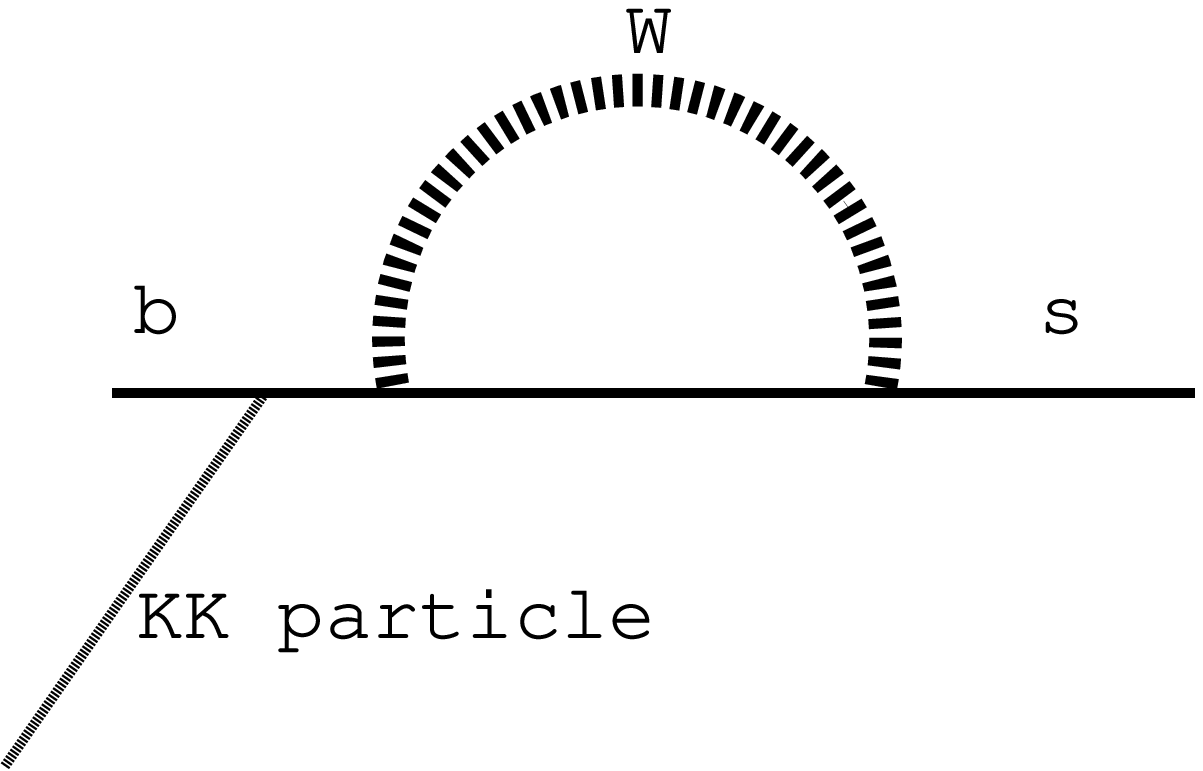}
\hskip 1.5cm
\epsfxsize=5cm\epsfysize=3.5cm
                    \epsfbox{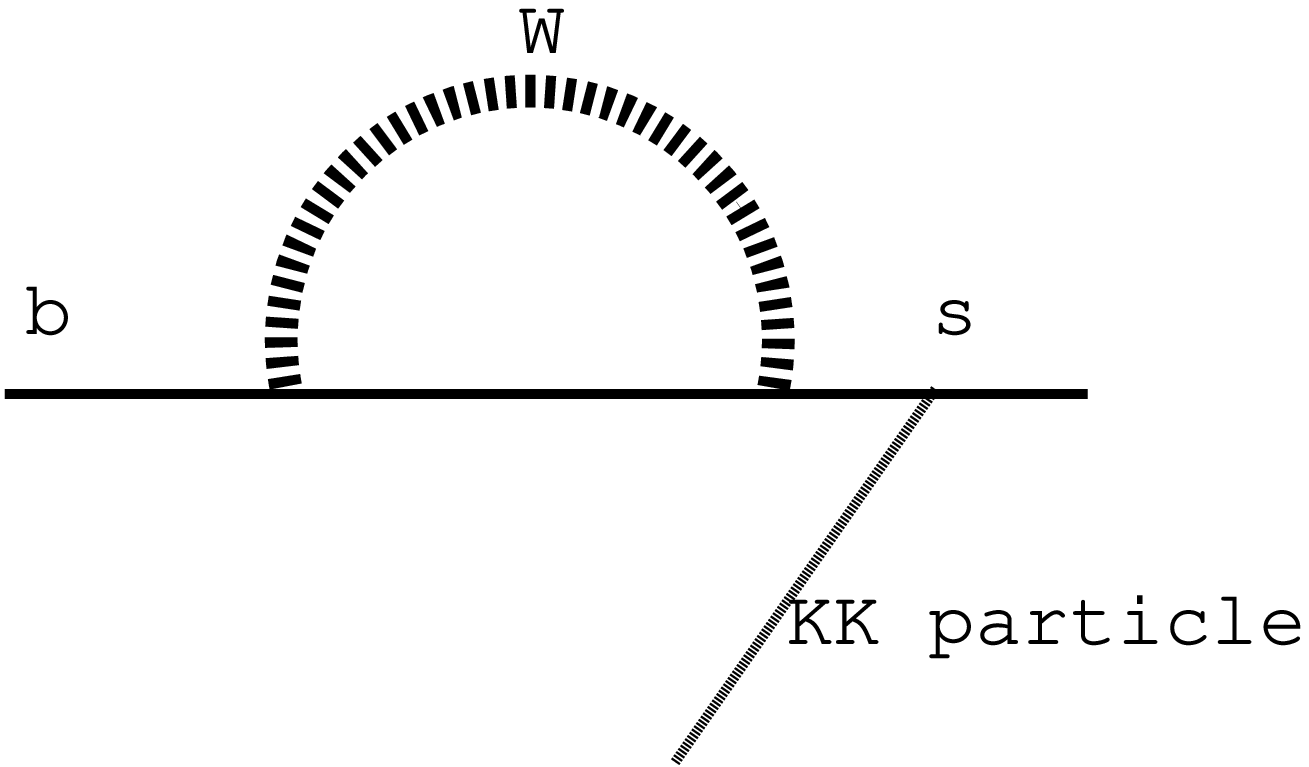}}
\caption{\em The diagrams contributing to the dilaton emission. We
label the diagrams as (a), (b), (c) and (d) moving clock-wise from top left.
For the graviton emission there are two more diagrams with the graviton
being hooked to either of the two ends of the loop.}
\end{figure}

\end{document}